\documentclass[aps,preprint,a4paper,floatfix]{revtex4-1}

\usepackage{graphicx,color}
\usepackage{amsmath,amssymb,amsfonts,amsthm,amscd,bm}
\usepackage{booktabs}
\usepackage{cancel}

\usepackage{amsmath}
\usepackage{amssymb}
\usepackage{amsfonts}
\usepackage{amsthm}
\usepackage{amscd}
\usepackage{bm}
\usepackage{graphicx}
\usepackage{booktabs}
\usepackage{listings}
\usepackage{xcolor,colortbl}
\usepackage{float}

\usepackage{amsthm}
\usepackage{float}
\usepackage{tikz}

\usepackage{amsmath}
\usepackage{amssymb}
\usepackage{amsfonts}
\usepackage{amsthm}
\usepackage{amscd}
\usepackage{bm}
\usepackage{graphicx}
\usepackage{booktabs}
\usepackage{listings}
\usepackage{xcolor,colortbl}
\usepackage{float}

\def\tilde{\widetilde}
\def\bar{\overline}
\def\hat{\widehat}
\def\*{\star}
\def\[{\left[}
\def\]{\right]}
\def\({\left(}      

\def\){\right)}

\def\frac#1#2{\dfrac{#1}{#2}}
\def\inv#1{\dfrac{1}{#1}}
\def\half{\tfrac{1}{2}}

\def\2pi{\hbox{$2\pi i$}}

\def\dsl{\raise.15ex\hbox{/}\kern-.57em\partial}
\def\Dsl{\,\raise.15ex\hbox{/}\mkern-.13.5mu D}
\def\2pi{\hbox{$2\pi i$}}

\def\dsl{\raise.15ex\hbox{/}\kern-.57em\partial}
\def\Dsl{\,\raise.15ex\hbox{/}\mkern-.13.5mu D}

\def\barray{\begin{eqnarray}}
\def\earray{\end{eqnarray}}
\def\beq{\begin{equation}}
\def\eeq{\end{equation}}

\def\n{\noindent}

\def\AA{\leavevmode\setbox0=\hbox{h}
\dimen0=\ht0 \advance\dimen0 by-1ex\rlap{\raise.67\dimen0\hbox{\char'27}}A}

\def\half{\tfrac{1}{2}}

\def\rhohat{\hat{\rho}}
\def\phat{\hat{p}}
\def\ptilde{\tilde{p}}

\def\rhobar{\bar{\rho}}
\def\rhoLambda{\rhobar_{\rm de}}

\begin{document}

\bibliographystyle{plainnat}

\title{{\Large {\bf Extensions of the Schwarzschild solution into regions of non-zero energy density and pressure}
 }} 

\author{
 Andr\'e  LeClair\footnote{andre.leclair@gmail.com}
}
\affiliation{Cornell University, Physics Department, Ithaca, NY 14853}

\begin{abstract}
\noindent
\noindent

We present solutions of the Einstein equations that extend the static  Schwarzschild solution in empty space into regions of non-zero energy density $\rho$ 
and radial pressure $p= w \rho$,  where $w$ is a constant equation of state parameter.  
For simplicity we focus mainly on solutions with constant $\rho$.  For $w=0$ we find solutions both with and without a singularity at the origin.   We propose that our explicit  non-singular solution with $w=-1$   describes the interior of a black hole,  which is a form of vacuum energy.  We verify  that its entropy is consistent with the 
Bekenstein-Hawking entropy,  however one needs to  assume the Hawking temperature.         We  further suggest that this idea can perhaps be  applied to the dark energy of the observable universe,   if one  views the latter  as arising from  black holes as  pockets of vacuum energy.
We estimate  the average  density of such a dark energy  to be $\rhoLambda \approx 10^{-30} g/{\rm cm}^3$.      
We also  present solutions with non-constant $\rho \propto 1/r^2$.

\end{abstract}

\maketitle

\section{Introduction}


The Schwarzschild solution  \cite{SchwarzschildRef} to Einstein's equations is based on  a zero stess-energy tensor $T_{\mu \nu}$.   It is static and spherically symmetric,  and the total mass is $M$.      
It was eventually understood  that it has an event horizon at the Schwarzschild radius $r_s = 2 MG/c^2$,  and  this initiated the  
development of the theory of black holes.    The solution has a  true singularity at the origin $r=0$.     The standard point of view is that anything that falls into the 
black hole beyond $r_s$ will eventually reach the singularity,   and perhaps quantum gravity effects can resolve this singularity.   It is not clear  that this is the only
possible resolution.           It would  thus be desirable to  discover a  classical resolution of the singularity within the context of General Relativity   which predicts  some properties  of the interior of 
black holes,  and this is the subject of this work.     

Theories for the internal structure of black holes may even be testable now that  gravitational waves 
from black hole mergers have been detected \cite{LIGO}.   These experimental results were the main motivation for our work.    There may be some signature of the internal structure  of a black hole  in the gravitational wave signal,   however we will
not address this complicated  aspect  here.   

As a simplifying approximation,  we introduce a length scale $r_0$ such that stress-energy tensor satisfies
\beq
\label{Tzero}
T_{\mu \nu} = 0  ~~{\rm for}  ~ r> r_0, ~~~~~T_{\mu\nu} \neq 0 ~~{\rm for} ~ r \leq r_0.
\eeq
Thus,  for $r>r_0$ the solution must be equal to the Schwarzschild solution with total mass $M$. 
In finding solutions,  one needs to match  the  Schwarzschild solution at 
$r=r_0$.   Although this is a rather straightforward approach,  it's not clear from the beginning that interesting exact solutions exist.   
However,  we will present many  such solutions,    some of which extend to the interior of a black hole.    

Let us summarize our main results.    Henceforth we set the speed of light $c=1$.  Planck's constant $\hbar \neq 0$ will only be relevant in Section \ref{blackhole} in connection with the 
Hawking temperature.         In the next section we present the Einstein equations in a form that we could  not find in the literature and are easier to solve.    We mainly focus on  solutions with constant non-zero energy density $\rho$ in the region $r<r_0$.   
In Section \ref{zerop} we find solutions with  zero radial pressure $p$, which is commonly attributed to matter.    We present solutions with and without singularities at $r=0$,  the latter being simpler.

In Section \ref{wsection} we consider non-zero radial pressure $p=w \rho$ where $w$ is a constant,  focusing  on non-singular solutions,     and  we present such solutions for arbitrary $w$.    The case $w=-1$ is especially interesting since the energy density and pressures are equivalent to vacuum energy,  i.e.  $T_{\mu\nu} = -\rho \, g_{\mu\nu}$.   
In Section \ref{blackhole} we study the black hole limit $r_0 = r_s$.    This solution has no singularity at the origin and does not have the peculiarity that the effective speed of light
vanishes inside the black hole as was the case  for $w=0$.        We propose that this is the correct  equation of state of the interior of a black hole.     As a check of this idea we show that
the classical  entropy of the black hole is consistent with the 
 Bekenstein-Hawking entropy if one assumes that the temperature is the Hawking temperature up to a factor of two.    The detailed nature of the matter inside the black hole that leads to $w=-1$  is not possible to address  based on this work. 
 
Having found these new solutions to the static Einstein equations, which are on solid  mathematical ground,   we proceed to try to apply  them to cosmology. This  necessarily should be viewed as  more  speculative since this is no longer a static situation.    Nevertheless,   our  attempts are are in line with standard approaches to cosmology.     For instance,   in the current universe,  which  has a large matter component (about $30\%$),   this matter content is treated as a fluid with zero pressure,  where the ``particles" are actual galaxies.   However the galaxies are treated as static sources in spite of the expansion of the universe,   in the sense that mergers of galaxies, accretion of galaxies, etc.  are  ignored,  and one still obtains reliable predictions.   
Black holes are also not static in that they inevitably grow,  but for  the above reasons we will treat them as static sources of gravitation as an approximation.    
Having stated these caveats,  in Section \ref{cosmoconstant} we propose that the {\it currently} observed non-zero dark energy  arises from the pockets of vacuum energy inside black holes.   This kind of dark energy is not a conventional cosmological constant because it changes with time due to the simple fact that the total energy density of black holes changes with time.    We  present this idea since the predictions are very good: 
we  roughly estimate the average energy density $\rhoLambda$ of such a proposed  dark energy  to be $\rhoLambda \approx 10^{-30} g/{\rm cm^3}$ which is surprisingly close to the 
measured value (``de" here refers to dark energy);    however this estimate is perhaps fortuitously too good (see the discussion below.)       

Finally in Section \ref{rhonotconstant} we present a solution with non-constant energy density $\rho \propto  1/r^2$.   However we do  not offer a potentially physical interpretation of this solution.


\section{The Einstein Equations}

 The general  spherically symmetric and static metric is defined by the line element 
 \beq
 \label{metric}
 ds^2 = g_{\mu\nu} dx^\mu dx^\nu = - e^{2a} \, dt^2 + e^{2b}\,  dr^2 + r^2 ( d\theta^2 + \sin^2 \theta  \,d\varphi^2 )
 \eeq
 
Further calculations  are  largely  based on Weinberg's comprehensive  book  \cite{Weinberg}.        The non-zero Christoffel symbols are  
 \barray
 \Gamma^t_{tr} &=& \Gamma^t_{rt} = a', ~~~~~\Gamma^r_{rr} = b' ,  ~~~~~~\Gamma^r_{tt} = a' e^{2(a-b)}
 \\
 \Gamma^\varphi_{r\varphi} &=&  \Gamma^\varphi_{\varphi r} = \Gamma^\theta_{r\theta} = \Gamma^\theta_{\theta r} = 1/r 
 \\
 \Gamma^r_{\theta\theta} &=&  \tfrac 1{\sin^2 \theta}\Gamma^r_{\varphi\varphi} = -r e^{-2b}
 \\
 \Gamma^\theta_{\varphi \varphi} &=& - \sin \theta \cos\theta, ~~~~~~ \Gamma^\varphi_{\varphi \theta}  = \cot \theta 
 \earray
 where $a'$ denotes the derivative with respect to $r$,  i.e. $a' = da/dr$, etc. 
The Einstein equations are 
\beq
\label{Einstein}
G_{\mu\nu} = R_{\mu \nu} - \half R\,  g_{\mu \nu} = 8 \pi G\,  T_{\mu \nu}
\eeq
where the non-zero components of $G_{\mu \nu}$ are 
\barray
\label{Einstein2}
G_{tt} &=& e^{2(a-b)} \( \frac{2 b'}{r} - \inv{r^2} \) + \frac{ e^{2a}}{r^2}
\\
G_{rr} &=& \frac{2 a'}{r} + \inv{r^2} \(1-e^{2b} \)
\\
\label{thetaphi}
G_{\theta \theta} &=& \tfrac{1}{\sin^2 \theta} G_{\varphi \varphi} = e^{-2b} r^2 \( a'' + a' \, ^2 - a' b' +(a'-b')/r \) 
\earray

We take the following form for the stress-energy tensor:
\barray
T_{tt} &=& \rho \, e^{2a}, ~~~~~T_{rr} = p\,  e^{2b}
\\
T_{\theta \theta} &=& \tfrac{1}{\sin^2 \theta} T_{\varphi \varphi} = p_\theta \, r^2 
\earray
where $\rho$ is the energy density and $p , p_\theta$ are pressures.  The metric factors $e^{2a}$ and $e^{2b}$ are necessary in the above equation:  for instance for vacuum energy 
$T_{\mu\nu} = - \rho \,  g_{\mu\nu} $.
   The pressures in $T_{\theta \theta}$ and $T_{\varphi \varphi}$ must be identical due to 
the  first eqality in \eqref{thetaphi}.   

It will be convenient to rescale $\rho$ and  the  pressures  as follows
\beq
\label{hats} 
\rhohat = 8 \pi G\, \rho, ~~~~ \phat = 8\pi G\, p, ~~~~~ \ptilde = 8 \pi G\, p_\theta 
\eeq
such that $\rhohat, \phat$ and $\ptilde$ all have dimensions of inverse length squared.  (If $\rho$ has units of mass per volume,  then reintroducing $c$, $\rhohat$ has   units of inverse time squared.)
The Einstein equations now read
\barray
\label{E1}
\frac{2b'}{r}  + \inv{r^2} \( e^{2b}-1 \) &=& \rhohat \,e^{2b} 
\\
\label{E2}
\frac{2a'}{r} - \inv{r^2} \( e^{2b} -1 \) &=& \phat \, e^{2b} 
\\
\label{E3}
a'' + a'^2 - a' b' +(a'-b')/r &=& \ptilde \,e^{2b} 
\earray

These equations are generally difficult to solve due to the second order equation \eqref{E3}.     However one can replace the latter    with a first order equation by differentiating \eqref{E2} and using \eqref{E1}.   One obtains 
a somewhat complicated equation,  however it  will turn out to be very useful:
\beq
\label{E3prime}
\ptilde = (4(r^2 \rhohat -1) )^{-1}  \[ r^2 \phat^2 (2rb'-1) + 2 \phat \( r^2 \rhohat(1+rb') + rb' -2 \) 
+ r \( 2 \rhohat ( b' + r^2 \phat' ) -2 \phat' -r\rhohat^2 \) \]
\eeq
  This equation can be understood as a statement of energy-momentum conservation since Bianchi identities ensure that $T_{\mu \nu}$ is covariantly   conserved  $D^\mu T_{\mu\nu} =0$  \cite{Weinberg}.  
  
  We will refer to $\phat$ as the radial pressure and  $\ptilde$ as the orbital pressure.     One  novelty of our work  is that the  form of the  above equations \eqref{E1}, \eqref{E2}, and  \eqref{E3prime} are easier to solve analytically compared to other equivalent versions in the literature.   In particular,  when $\rhohat$ and $\phat$ are specified,  then equations \eqref{E1} and \eqref{E2} already determine  solutions,  up to constants of integration and  the boundary condition at $r=r_0$;   then the orbital pressure $\ptilde$ is determined by 
  \eqref{E3prime}.   
  
\section{The Schwarzschild Solution}

To warm up, it is instructive to reproduce the known Schwarzschild solution from the above equations.  Here,  we assume a central point mass $M$ at $r=0
$,   and for $r>0$ we have  $\rhohat = \phat = \ptilde =0$,  namely $r_0 = 0$.         Using \eqref{E1} and \eqref{E2} 
one finds 
\beq
\label{SchwarzSol} 
2b = - \log \( 1-r_s /r \),    ~~~~~  2a = \alpha + \log \(1-r_s /r \)
\eeq
where $r_s$ and $\alpha$ are constants of integration.    Imposing  that $g_{\mu \nu}$  approaches the Minkowski empty space metric as $r\to \infty$,    and using $g_{tt} = - e^{2a}  \to -1 - 2 \Phi(r)$ where $\Phi(r) = - MG/r$ is the Newtonian potential \cite{Weinberg},  one finds $\alpha =0$ and 
\beq
\label{SchwarzSol2}  
e^{2a} = \( 1- \frac{r_s}{r} \) , ~~~~~ e^{2b} = \(1 - \frac{r_s}{r} \)^{-1}
\eeq
where $r_s$ is the Schwarzschild radius
\beq
\label{rs} 
r_s \equiv  2 M G
\eeq
It is important to note that  the above solution  depends on determining the constant of integration and that \eqref{E3} is automatically satisfied due to the equivalent equation \eqref{E3prime}.  

\section{Solutions  with  constant non-zero  energy density,  and zero  radial pressure }

\label{zerop}

\subsection{General Solutions ignoring the boundary conditions.}

In this section we consider solutions with constant energy density $\rho$,   and  with zero radial pressure,  i.e. $p=0$,  which is commonly associated with cold matter.      
We will  ignore  the  boundary condition \eqref{Tzero} at $r=r_0$  for the remainder of this sub-section, but will impose it later.     

Since $\rhohat$ has units of inverse length squared,  let us define the scale $\ell$ 
\beq
\label{elldef} 
\rhohat \equiv  1/ \ell^2  , ~~~~ \phat =0
\eeq
where $\ell$  has units of length and is constant in $r$.    
The solution to \eqref{E1} and \eqref{E2} 
is 
\barray
\label{beq}
2b &=& - \log \( 1 - \frac{r^2}{3 \ell^2}   -  \frac{\beta \ell}r \) 
\\
\label{aeq} 
2 a &=& \alpha - \log( r/\ell)  + \sum_{x-{\rm roots}}\(  \frac{ \log (x/\ell - r/\ell)}{1-x^2/\ell^2 } \)
\earray
where $\alpha$ and $\beta$ are constants of integration,  and 
 $x$-roots  are the three roots of the cubic  algebraic equation 
\beq
\label{xroots} 
3 \beta - 3 (x/\ell) + (x/\ell)^3 = 0
\eeq

\subsection{Solution for all $r$  with no singularity at $r=0$ 
 and its Black Hole limit.}

Let us first impose that there is no singularity in $b(r)$  at $r=0$.   Then $\beta =0$ and the $x$-roots are simply 
\beq
\label{xroots0}
x/\ell = 0, ~~~~ x/\ell = \pm \sqrt{3}
\eeq
By redefining the constant $\alpha$ one finds the solution 
\beq
\label{rhosol}
e^{2a} = \frac{ C^2}{\sqrt{1-\frac{r^2}{3 \ell^2}}} \, ,  ~~~~e^{2b} = \( 1-\frac{r^2}{3 \ell^2} \)^{-1}
\eeq
where  $C$  is a real  constant.   

Next we require that the above solution matches the Schwarzschild solution at $r=r_0$:
\barray
\label{match} 
\(1-\frac{r_s}{r_0} \)^{-1}  = \( 1 - \frac{r_0^2}{3 \ell^2} \)^{-1}
\\
1-\frac{r_s}{r_0} = C^2 \(  1 - \frac{r_0^2}{3 \ell^2 } \)^{-1/2} 
\earray
The first equation is easily interpreted since it simply implies 
\beq
\label{Mrho}
\frac{r_s}{r_0} = \frac{r_0^2} {3 \ell^2} , ~~~~~ \Longrightarrow ~~~~\frac{4}{3} \pi r_0^3 \rho = M
\eeq
Multiplying the two equations,  one concludes
\beq
\label{C2}
C^2 = \( 1 - \frac{r_0^2}{3 \ell^2} \)^{3/2}
\eeq
Requiring $C^2$ to be real implies $r_0^2/2 \ell^2 \leq 1$,  and together with \eqref{Mrho}
\beq
\label{rsro}
r_s \leq r_0.    
\eeq
Finally,   using \eqref{E3prime}  one finds the orbital pressure has the simple expression: 
\beq
\label{porbital}
\ptilde = \inv{4 \ell^2}  \(\frac{3 \ell^2}{r^2} -1\)^{-1}
\eeq

Recall there is no matter for $r>r_0$ where the solution is the usual Schwarzschild solution.    The equation \eqref{rsro}  implies that the Schwarzschild radius is 
generally 
{\it inside} the region with a distribution of matter where $r<r_0$,  so that our solution does not represent the interior of a black hole,  except  possibly for  $r_s=r_0$,  which will be considered below.     It is important to note  that our solution is not singular at $r=0$.

\bigskip

\n 
{\bf Black Hole limit?}
Consider the limiting case $r_s = r_0$,  which is in principle allowed by \eqref{rsro},  where $r_0$ approaches $r_s$ from above.      In this case,  all the matter is inside the Schwarzschild radius,   and our solution arguably extends the 
Schwarzschild solution into the interior of the event horizon of a black hole,  which is rather intriguing.     
At $r=0$ the metric is 
\beq
\label{metric0}
ds^2 = - C^2 dt^2 +  dr^2 
\eeq
where $C$ can be interpreted as an effective speed of light.   However \eqref{Mrho} and \eqref{C2} imply that the effective speed of light $C=0$.    
This would seem to be consistent with the fact that light cannot escape the  event horizon,   since it is slowed down to zero speed  everywhere inside,  namely,
$C=0$ for all $r<r_s$.   It is as if time has stopped so no longer exists.       Also note that the orbital pressure 
$\ptilde =0$ at $r=0$.     However it is not entirely clear that $C=0$ is physically sensible,  and we think it actually is not;   we will thus not deliberate further on this limit  in this paper.     Fortunately in Section \ref{wsection} we will find a 
more physically appealing  solution with a different equation of state,  namely $p= w\rho$,  with $w=-1$,   as for vacuum energy.

\subsection{Solutions with $\beta \neq 0$.}

\def\bt{\tilde{\beta}}

Let us now turn to solutions with a singularity at the origin where $\beta \neq 0$.     The solutions are more complicated,   but nevertheless have analytic expressions.    

Define
\beq
\label{btdef}
\bt =  \frac{3 \beta}{2} \( 1- \sqrt{1 -\frac{4}{9\beta^2}  } \)  
\eeq 
The roots to \eqref{xroots} are then 
\barray
x_1/\ell  &=&   e^{i\pi/3} \, \bt^{1/3}  + e^{-i\pi/3}\, \bt^{-1/3} 
\\
x_2/\ell  &=& - \bt^{1/3} - \bt^{-1/3} 
\\
x_3/\ell  &=&  -  e^{2 \pi i/3} \,\bt^{1/3} - e^{-2\pi i/3} \, \bt^{-1/3} 
     \earray
We express the solution in the above form in order to properly keep track of branches.   

In order for the metric to be real,   we require that $\beta$ is real.  
There are two cases to consider:

\bigskip
\n
{\bf Case 1.} ~~   Here  $|\beta| > 2/3$ and $\bt$ is real.    In general the 
$x_i$ are complex,   however we only require that $a(t)$ is real,  which is compatible with complex $x_i$.   
For instance when  $0<\bt<1$,  then it  turns out that $x_3 = x_1^*$ and $x_2$ is real,  which implies that $a(r)$ in \eqref{aeq} is real. 
On the other hand when  $\bt<0$,  $x_2 = x_1^*$ and $x_3$ is real  so that again $a(r)$ is real.   
  Incidentally,  it  turns out that if $0<\bt<1$ is rational,  then so is $\beta$.    
  For instance  $\bt = 9/10$  corresponds to $\beta = 181/270$.  
  
  \bigskip
\n 
{\bf Case 2.} ~~   Here $|\beta| <2/3$,   and
\beq
\label{btphase}
\bt = 3 \beta/2 - i  \sqrt{1-9\beta^2/ 4} 
\eeq
One sees that $|\bt| = 1$ such that $\bt$ is a pure phase.    Let us parameterize it as
\beq
\label{btphase2}
\bt = e^{-3 \pi  i \gamma },   ~~~~~~~~~~\gamma \equiv  \inv{3 \pi} \arccos (3 \beta /2)
\eeq
A nice feature of this case is that the $x$-roots  are real:
\beq
\label{xrootscos}
\{ x_1, x_2, x_3 \} /\ell  = \{  2  \cos \( \pi(\gamma-\tfrac{1}{3}  ) \) ,  -2 \cos (\pi \gamma),  ~ 2 \cos \( \pi (\gamma+\tfrac{1}{3})\)  \}  
\eeq
These roots satisfy 
\beq
\label{sumroots} 
x_1 + x_2 + x_3 =0
\eeq
The case $\beta =0$ of the last section corresponds to $\gamma=1/6$ and $\bt = -i$. 
   
Cases 1 and 2 are separated by $\beta = 2/3$,   where $\gamma =0$ and $\{ x_1, x_2, x_3 \}/\ell  = \{1, -2, 1\}$.   The roots $x/\ell=1$ lead to singularities in the
expression \eqref{aeq} for $a(r)$,  and won't be further considered here.

 \bigskip
 
\def\bigAbsl{\Big |}
\def\bigAbsr{\Bigr |}

Rather than  attempt to solve all such cases,  for illustrative and simplifying  purposes,  we will limit ourselves  to Case 2  where  all $x$-roots are real.    
Returning to \eqref{aeq},     for each $x_i$ one has $\log( (x_i- r)/\ell ) = \log |(x_i-r)/\ell |  ~~{\rm or}  ~~   \log  |(x_i-r)/\ell |  + i \pi $ depending on the sign of $(x_i-r)$.   If $r $ is such that the 
$i\pi$ is required,   then  this just leads to a constant that can be absorbed into the constant $\alpha$.     Consequently,  defining
\beq
\label{nu}
\nu_i  = \inv{\(1-x_i^2 /\ell^2 \)}
\eeq
one finds the solution
\barray
\label{Sol1}
e^{2b} &=& \(  1 - \frac{r^2}{3 \ell^2} - \frac{\beta \ell}{r} \)^{-1} 
\\
\label{Sol2}   
e^{2a} &=&  C^2 ~ \( \frac{\ell}{r} \)\,  \prod_{i=1}^3 \bigAbsl 1-  \frac{r}{x_i} \bigAbsr^{\nu_i}
\earray
Above,   $C^2 $ is required to be a real constant which implicitly depends  on $\beta$.    
Matching the above solution to the Schwarzschild solution at $r=r_0$,   one obtains
\barray
\label{rsrobeta}
\frac{r_s}{r_0} &=& \frac{r_0^2}{3 \ell^2} + \frac{\beta \ell}{r_0} 
\\
\label{C2o}
C^2 &=& \frac{r_0}{\ell} \( 1- \frac{r_s}{r_0} \) \, \prod_{i=1}^3 \bigAbsl  1- \frac{r_0}{x_i}  \bigAbsr^{-\nu_i} 
\earray

\section{Solutions with non zero radial pressure}

\label{wsection}

Here we consider 
\beq
\label{pw}
\rhohat = 1/ \ell^2 , ~~~~~ \phat  = w \, \rhohat
\eeq
where $\ell$ and $w$ and constants.    In cosmology,  for matter,  radiation,  and vacuum energy,   $w= 0, 1/3, -1$ respectively,  and these are still interesting cases in 
our context.

\subsection{ General $w$ } 

The solution of the last section still applies with modified exponents $\nu_i$.      Namely \eqref{Sol1} and \eqref{Sol2} still apply,   where the $x_i$ satisfy the same 
cubic equation \eqref{xroots},      however now
\beq
\label{nuw} 
\nu_i = \frac{ 1 + w\,  x_i^2/\ell^2}{1- x_i^2/ \ell^2}
\eeq
Matching to the Schwarzschild solution at $r=r_0$,  one again obtains \eqref{rsrobeta} and \eqref{C2o} with these modified exponents $\nu_i$.

For physical reasons,  and also for simplicity,   let us consider non-singular solutions in $e^{2b}$ where $\beta =0$.    Then, as before,   $\{ x_i \} /\ell= \{0, \pm \sqrt{3} \}$.  One sees that the 
$x_i =0$ root  just  cancels the $\log r$'s in \eqref{aeq},  such that the $1/r$ factor in  \eqref{Sol2} is cancelled by the $x_i = 0$ root.      
One obtains
\beq
\label{e2anu}
e^{2b} = \(  1 - \frac{r^2}{3 \ell^2}  \)^{-1} , ~~~~~
e^{2a} = C^2  ~ \bigAbsl 1-\frac{r^2}{3\ell^2} \bigAbsr^{\nu},   ~~~~~~~~~~~~~\nu = -(1+ 3 w)/2 
\eeq
Matching to the Schwarzschild solution at $r=r_0$,  one finds
\beq
\label{wMatch}
\frac{r_s}{r_0} = \frac{r_0^2}{3 \ell^2} ,~~~~~~C^2 =  \( 1- \frac{r_s}{r_0} \)  \bigAbsl  1- \frac{r_s}{r_0} \bigAbsr^{- \nu} 
\eeq

\subsection{The special case $w=-1$}

Unless $w=-1$,  $C^2$ equals $0$ or $\infty$ as $r_0 \to r_s$.    When $w=-1$,  $\nu =1$,  
  and the above solution simplifies considerably.     
Remarkably,   from \eqref{E3prime} one obtains the non-trivial result  from this  complicated expression that $\ptilde = -1/\ell^2$.   Thus when $w = -1$,  all pressures are  entirely consistent with vacuum energy,  i.e.  $T_{\mu\nu} = - \rho \, g_{\mu\nu}$:
\beq
\label{cconstant}
\{ \rhohat, \phat , \ptilde \} = \{ 1, -1, -1 \}/\ell^2
\eeq 
In the next section,  we will apply this solution to the interior of a black hole.

\section{The interior of a black hole as vacuum energy}

\label{blackhole}

\subsection{The solution inside the event horizon}

Consider the non singular solution of the last section  ($\beta =0$)  with $w=-1$ where $\nu = 1$.     Recall that by construction for $r>r_0$ the solution is the Schwarzschild one.   
Let $r_0 \to r_s$,   approaching the limit from above.   In this limit,   all the matter is inside the event horizon at $r_s$ and can be interpreted as a black hole of mass $M$.    
It turns out that in this limit $C^2=1$,  which is more physically sensible than $C^2 =0$ in Section \ref{zerop} where $w=0$.     
In fact,  unless $w=-1$,   $C^2$ is equal to either $0$ or $\infty$,   so that $w=-1$ is the only physically sensible  possibility.      Inside the black hole $r<r_s$ the solution is  quite simple.
\beq
\label{BHsolution}
e^{2a}   =  1-\frac{r^2}{3 \ell^2}, ~~~~~~ e^{2b} = \(1- \frac{r^2}{3 \ell^2} \)^{-1}  ~~~~~~( r<r_s )
\eeq
where $r_s/r_0 = r_0^2/3 \ell^2$,  which just implies \eqref{Mrho}.      The original black hole singularity at $r=0$  of the Schwarzschild solution no longer exists.    Furthermore,    inside the black hole,  
the energy and pressures are interpreted as vacuum energy due to \eqref{cconstant}.

\subsection{Black Hole Entropy}

Basic laws of thermodynamics, with zero chemical potential,   imply   
\beq
\label{FirstLaw}
T \, dS = dU + p \, dV
\eeq
Here $U$ is the internal energy,  so that $U= \rho\, V$.    
One has $dU = \rho \, dV + d\rho \,V$ and the $\rho \, dV$ term is cancelled  when $p= - \rho$.  
Thus 
$dS/d\rho = V/T$,   which implies
$S= V( \rho  - \rho_0)/T$, where $\rho V = M$ is the mass of the black hole and $\rho_0$ is a constant of integration.    Now,  $M/T$ is a constant,  and 
if  $M=0$,  then nothing exists,   and the entopy $S$ must be  zero.   Thus we take the  integraton constant to be proportional to $\rho$.   
\beq
\label{SVT2}
S =  \kappa \,  \frac{M}{T} = \kappa\,  \frac{r_s}{2 G T}.  
\eeq
for some constant $\kappa$,  which we cannot predict.

Thus far,  our analysis has been purely classical with $\hbar =0$.       Although we cannot justify the following  based on this  classical analysis,  let us nevertheless 
identify the temperature $T= T_H$ where $T_H$ is the Hawking temperature \cite{Hawking} 
\beq
\label{TH} 
T_H = \frac{\hbar}{k_B} \,  \inv{8 \pi G M }
\eeq
with  $k_B$  equal to  Boltzmann's constant.   Identifying the area $A=4 \pi r_s^2$,  one finds an  entropy which is proportional to the area \cite{Bekenstein,Hawking}.
For $\kappa = 1/2$,  which corresponds to $\rho_0 = \rho/2$,      one finds the Bekenstain-Hawking entropy 
\beq
\label{BH2}
S = \frac{k_B}{4 \ell_p^2}  \, \, A 
\eeq
where  $\ell_p = \sqrt{\hbar G}$ is the Planck length.    One point of view is that one can chose the constant of integration  $\rho_0$ such that $\kappa=1/2$ in order to 
match with the Bekenstein-Hawking entropy;  however as stated,  we could not present arguments to justify this based on our classical analysis.     Furthermore it is not possible to predict the nature of the matter inside the black hole beyond its equation of state $w=-1$.

\section{Could the Dark Energy of the observable universe  originate from the vacuum energy  inside  black holes?
}

\label{cosmoconstant}

\def\bh{{\rm bh}}
\def\g{{\rm g}}
\def\rhobar{{\bar{\rho}}}

In the last section,   we have proposed that the interior of a black hole consists of vacuum energy.      Just as the matter content of the universe is treated as a fluid,  where the ``particles" are individual  galaxies,   black holes can be considered as pockets of vacuum energy,   and on average  constitute  a total vacuum energy density   $\rhoLambda $.  One should distinguish between the local effect of black holes,   which are classically point masses beyond the Schwarzschild radius,   verses their global cosmological effects.   We suggest that 
the latter can  perhaps be interpreted as  dark energy.      In the following,  we only make rough estimates.  

\subsection{The density of  the proposed dark energy for the observable universe}

\def\rhobar{\bar{\rho}}

Let $\rho_{\rm de}$ denote the total vacuum energy of the universe   due to black holes {\it at the current time}.    
Then
\beq
\label{BHtotal}
\rho_{\rm de}  = \frac{M_{\bh-{\rm  total}}}{V_{\rm total}}
\eeq
where  $M_{\bh - {\rm total}}$ is the  sum total of the masses of all black holes in the universe and $V_{\rm total}$ is the total volume of the universe.   
We can give a rough estimate of $\rho_{\rm de}$  as follows.   
The average $\rhoLambda$ is 
\beq
\label{rhoLambda}
\rhoLambda   = \frac{\bar{V}_{\bh}}{V_{\rm total}} \, N_\g \,  \bar{N}_{\bh}\,  \bar{\rho}_{\bh}  = \frac{N_\g \, \bar{N}_{\bh} \,  \bar{M}_{\bh} }{V_{\rm total} }
\eeq
where $\bar{V} _{\bh}$ is the average volume of a black hole,     $N_\g$ is the total number of galaxies,   $\bar{N}_{\bh}$ is the average number of black holes per galaxy,  $\bar{\rho}_\bh$ is the average density of a black hole, 
and $\bar{M}_{\bh}$ is the average mass of a black hole.      
If $\rho_{\rm total}$ is the total energy density  of the universe, 
then 
$\rhoLambda / {\rho_{\rm total}} =  M_{\bh-{\rm  total}}/{M_{\rm total}}  < 1$  where  $M_{\rm total}$ is the total mass of the universe.   
 It needs to be emphasized  that $\rhoLambda$  in \eqref{rhoLambda} is not constant in the evolution of the universe, {\it and is thus not an ordinary cosmological constant}.

The above formula \eqref{rhoLambda} for    $\rhoLambda$    is compatible with the measured dark energy density   $\rho_{\rm de}  = 0.7 \times 10^{-29} \, g/{\rm cm}^3$  ~\cite{CC1,CC2},    for some  reasonable estimates of the
parameters.       For instance,  let us take  the known estimates  $N_\g = 10^{11}$,    $\bar{N}_{\bh} =10^8$ as in our own galaxy,   and the estimate 
$V_{\rm total} = 4 \times  10^{86} \, {\rm cm}^3$.    The average black hole mass $\bar{M}_\bh $  is harder to estimate since it can range from $10 M_\odot$ to $10^9 M_\odot$.   
Given this  situation,  it makes sense to take the geometric mean of these extreme limits. Thus we take 
$\bar{M}_\bh = 10^5 M_\odot$.      Then  \eqref{rhoLambda} gives $\rhoLambda \approx 10^{-30} \, g/{\rm cm}^3$,   which is surprisingly close to the measured value.   However our estimates were rough approximations,  so this apparently excellent agreement is likely to be fortuitous.   
In particular,   our estimate for $\bar{M}_\bh $ is perhaps over estimated,  unless there are many as yet undetected intermediate mass black holes.       
  In any case,   more accurate estimates of $\rhoLambda$ along the above lines are likely to give the same value within an order of magnitude or so.

\section{ Some solutions with non-constant  energy  density}

\label{rhonotconstant}

\def\lhat{\hat{\ell}}

For the previous sections,  we considered constant energy density $\rho$.    In this section we consider the case
\beq
\label{rhor}
\rhohat  = \frac{\sigma}{r^2},  ~~~~
 \phat  =0   
 \eeq
where $\sigma$  is a constant.   
One can find an exact solution for any $\sigma$, which we do not present here.    The case $\sigma=1$ is particularly simple:
\beq
e^{2b} =  
 \frac{c_1  r} \lhat , 
~~~~~
e^{2a} = \(  \frac{c_2 \lhat }{r} \) ~ e^{c_1  r/\lhat}
\eeq
where $c_{1,2}$ are constants of integration,  and $\lhat$ is a length scale.      
  Matching to the Schwarzschild solution,  one finds
\beq
\label{c1c2} 
c_1 = \frac{\lhat}{r_0}\(  1 - \frac{r_s}{r_0}  \)^{-1}  , ~~~~~c_2 =  \(\frac{r_0}\lhat  -\frac{ r_s}\lhat \) \, e^{-c_1 r_0/\lhat}  
\eeq
Note that this solution is unavoidably singular at $r=0$,  and we don't have much more to say about it here.    

\section{Concluding Remarks}

We already summarized our results and proposals in the Introduction,  so let us just remark on some open questions,  of which there are many.  

\bigskip
\n
$\star$    We proposed that the interior of a black hole is vacuum energy.   
What is interesting about this proposal is that it  has nothing to do  with Quantum Mechanics nor Quantum Gravity.     We only introduced $\hbar$ in order to 
compare with the Bekenstein-Hawking entropy.    Is there indeed a   purely classical resolution of the original black hole singularity as we proposed?      A related question is whether the 
temperature $T$  in the black hole entropy formula  \eqref{SVT2} is necessarily the Hawking temperature $T_H$ which does depend on $\hbar$.  For $\kappa =1/2$,  or equivalently,  $T=2 T_H$,  our proposed entropy agrees with the Bekenstein-Hawking entropy,   however we could not justify this based on our classical analysis.  
     String theory models suggest that 
the temperature should indeed  equal $T_H$ \cite{StromingerVafa}.      

\bigskip
\n
$\star$  We have proposed that the vacuum energy inside black holes could perhaps explain the current dark energy density of the observable universe,  which is a late time inflation. 
The vacuum energy density due to black holes is not constant in time.   
Could the same idea be applied to the very early universe,   where the origin of inflation is a single primitive black hole?   Certainly the energy density in the early universe was large enough for such a black hole to momentarily exist.        We  have some preliminary  results in this 
direction \cite{LeClairInflation}.

\bigskip
\n
$\star$  Is there a physical application of  the singular solutions for zero radial pressure presented in Section \ref{zerop},   in particular those for Case 1, which were not studied in much detail?

\bigskip
\n
$\star$   Are there physically sensible solutions for the interior of a black hole where the matter is deeply inside, i.e. $r_0< r_s$?   Based on \eqref{rsrobeta} this may be possible for non-zero $\beta$ where $r_0^2/3 \ell^2 > 1 - \beta \ell/r_0$.

\bigskip

\n
$\star$   Since the measurement of gravitational waves from black hole mergers is now possible \cite{LIGO},   is it feasible  to detect any potential internal structure of a  black hole,
as of the kind  proposed in this work?

\bigskip\bigskip

\bigskip 
\noindent{{\bf Note added:}   After completing this article we were informed of previous works on non-singular black holes with  negative pressure from a different approach.  
In particular we wish to cite  the works 
\cite{Hayward,Mottola,Ramy1,Ramy2} and references therein.        We thank R. Brustein for pointing this out and for discussions.



\newpage

\begin{thebibliography}{99}

\bibitem{SchwarzschildRef}  K. Schwarzschild,  
{\it  Uber das Gravitationsfeld eines Massenpunktesnach der Einsteinschen Theorie}, 
Sitzungsberichte der Königlich Preussischen Akademie der Wissenschaften (1916).


\bibitem{LIGO}  LIGO and Virgo scientific collaborations,   	Phys. Rev. Lett.~ {\bf  116}, 061102 (2016).

\bibitem{Weinberg}  S. Weinberg,
{\it Gravitation and Cosmology},  John Wiley and Sons (1972).
 

\bibitem{VelocityCurves}   V. Rubin,  N. Thonnard,  and W. K. Ford Jr.,  
Astrophysical Journal {\bf  238}  (1980). 


\bibitem{Bekenstein}   J. Bekenstein,  
{\it  Black Holes and Entropy}, 
Phys. Rev. D. {\bf 7 }  (1973).  

\bibitem{Hawking}  S. Hawking,  
{\it  Particle creation by Black Holes}, 
 Comm.  Math. Phys. {\bf 43}   (1975).
 
\bibitem{CC1}  
A.G.  Riess et.al., 
{\it
Observational evidence from supernovae for an accelerating universe and a cosmological constant}, 
Astronomical Journal {\bf 116}   (1998).   



\bibitem{CC2}  Planck Collaboration,  {\it Planck 2015 results I. Overview of products and scientific results},
 Astronomy and  Astrophysics  {\bf 594}   (2016) . 



\bibitem{LeClairInflation}   A.  LeClair, 
{\it  Early inflation based on a single  aboriginal black hole},   arXiv:1907.01407 [gr-qc].  

\bibitem{StromingerVafa}  
A. Strominger and C. Vafa, 
{\it 
Microscopic origin of the Bekenstein-Hawking entropy}, 
Phys. Lett. B (1996).   


 \bibitem{Hayward} 
 S. A. Hayward, 
 {\it Formation and evaporation of regular black holes}, Phys. Rev. Lett. {\bf 96}, 031103 (2006).

 \bibitem{Mottola}
 P. O. Mazur and E. Mottola, 
 {\it Surface tension and negative pressure interior of a non-singular
black hole}, 
 Class. Quant. Grav. {\bf 32}, no. 21, 215024 (2015).
 
 \bibitem{Ramy1}
 R. Brustein and A.J.M  Medved,
 {\it Resisting collapse: How matter inside a black hole can withstand gravity}, 
 Phys. Rev. D {\bf 99} 064019 (2019).   
 
 \bibitem{Ramy2}
 R. Brustein and A.J.M  Medved,
 {\it A maximal-entropy initial state of the Universe as a microscopic description of inflation}, 
 arXiv:1906.00989. 


\end{thebibliography}
\end{document}